\documentclass[12pt]{iopart}

\usepackage{graphicx}

\begin{document}
\title[Kolkata Paise Restaurant Problem]{Statistics of the  Kolkata Paise Restaurant Problem}
\author{Asim Ghosh$^1$, Arnab Chatterjee$^2$, 
Manipushpak Mitra$^3$  and Bikas K Chakrabarti$^{1,3}$}
\address{$^1$ Theoretical Condensed Matter Physics Division and Centre for Applied Mathematics \& Computational Science,
Saha Institute of Nuclear Physics, 1/AF Bidhannagar, Kolkata 700 064, India}
\address{$^2$ Condensed Matter and
Statistical Physics Section,
The Abdus Salam International Centre for Theoretical Physics,
Strada Costiera 11, Trieste I-34014, Italy}
\address{$^3$ Economic Research Unit, Indian Statistical Institute, 
203 Barrackpore Trunk Road, Kolkata 700108, India}
\eads{asim.ghosh@saha.ac.in, achatter@ictp.it, mmitra@isical.ac.in \& bikask.chakrabarti@saha.ac.in}
\begin{abstract}
\noindent We study the dynamics of a few stochastic learning strategies for the  ``Kolkata Paise Restaurant'' problem, where $N$ agents choose  among $N$ equally priced but differently ranked restaurants every evening such that each agent tries get to dinner in the best restaurant (each serving only one customer and the rest arriving there going without dinner that evening). We consider the learning strategies to be  similar for all  the agents and assume that each follow the same probabilistic or stochastic  strategy dependent on the information of the past successes in the game.  We show that some ``naive'' strategies lead to much better utilization of the services than some relatively ``smarter'' strategies. We also show that the service utilization fraction as high  as $0.80$ can result for a stochastic strategy, where each agent sticks to his past choice (independent of success achieved or not; with probability decreasing inversely in the past crowd size). The numerical results for utilization fraction of the services in some limiting cases are analytically examined.

\end{abstract}
\submitto{\NJP}
\maketitle

\section{Introduction}
The Kolkata Paise Restaurant (KPR) problem \cite{anindya,mathematica,proc} is a repeated game, played between a large number $N$ of agents having no interaction amongst themselves. In KPR problem,  prospective customers (agents) choose from $N$ restaurants each evening simultaneously (in parallel decision mode); $N$ is fixed. Each restaurant has the same price for a meal but a different rank (agreed upon by all  customers) and can serve only one customer any evening. Information regarding the customer distributions for earlier evenings is available to everyone. Each customer's objective is to go to the restaurant with the highest possible rank while avoiding the crowd so as to be able to get dinner there. If more than one customer arrives at any restaurant on any evening, one of them is randomly chosen (each of them are anonymously treated) and is served. The rest do not get dinner that evening.

In Kolkata, there were very cheap and fixed rate ``Paise Restaurants"  that were popular among the daily laborers in the city. During lunch hours, the laborers used to walk (to save the transport costs) to one of these restaurants and would miss lunch if they got to a restaurant where there were too many customers. Walking down to the next restaurant would mean failing to report back to work on time! Paise is the smallest Indian coin and there were indeed some well-known rankings of these restaurants, as some of them would offer tastier items compared to the others. A more general example of such a problem would be when the society provides hospitals (and beds) in every locality but the local patients go to hospitals of better rank (commonly perceived) elsewhere, thereby  competing with the local patients of those hospitals.   Unavailability of treatment in time may be considered as lack of the service for those people and consequently as (social) wastage of service by those unattended hospitals.

A social planner's (or dictator's) solution to the KPR problem is the following: the planner (or dictator's) asks everyone to form a que and then assigns each one a   restaurant  with rank matching the sequence of the person in the que on the first evening. Then each person is told to go to the next ranked restaurant  in the following evening (for the person in the last ranked restaurant this means going to the first ranked restaurant). This shift process than continuous for successive evenings. Call this  solution the {\it fair social norm}. This is clearly one of the most efficient solution (with utilization fraction $\bar f$ of the services by the restaurants equal to unity) and the system arrives at this this solution immediately (from the first evening itself). However, in reality this cannot be the true solution of the KPR problem,  where each agent decides on his own (in parallel or democratically) every evening, based on complete information about past events. In this game, the customers try to evolve a learning strategy to eventually get dinners at the best possible ranked restaurant, avoiding the crowd. It is seen, the evolution these strategies take considerable time to converge and even then the eventual utilization fraction $\bar{f}$ is far below unity. The KPR problem have some basic features similar to the minority game problem \cite{arthur1994,challet2005} in that diversity is encourage (compared to herding behavior) in both, while it differs from (two-choice)  minority games in terms of the macroscopic size of the choices. 

As already shown in ref \cite{anindya}, a simple random-choice algorithm, if adapted by all the agents, can lead to a reasonable value of utilization fraction ($\bar{f}\simeq0.63$). Compared to this, several seemingly  ``more intelligent'' stochastic algorithms lead  to lower utilization of the services.  Ref. \cite{proc} studied a few more such ``smarter'' algorithms, having several attractive features (including analytical estimate possibilities), but still failing to improve the overall utilization fraction beyond its random choice value. Here we develop a stochastic strategy, which maintains  a naive tendency (probability decreasing with past crowd size) to stick to any agent's own  past choice (successful or not), leading to a maximum, so far,  value  of the utilization fraction $\bar f$ ($\simeq0.80$) in the KPR problem. We also estimate here analytically the $\bar f$ values for several of such strategies.

\section{Stochastic learning strategies}
 Let the symmetric stochastic strategy chosen by each agent be such that at any time $t$, the probability $p_k(t)$ to arrive at the $k$-th ranked restaurant is given by
\begin{equation}
p_k(t) =\frac{1}{z}\left[k^{\alpha}\exp\left(-\frac{n_k(t-1)}{T}\right)\right],\hspace{.1in} z=\sum_{k=1}^N\left[k^{\alpha}\exp\left(-\frac{n_k(t-1)}{T}\right)\right],\label{generalstoch}
\end{equation}
where $n_k(t)$ denotes the number of agents arriving at the $k$-th ranked restaurant in period $t$, $T>0$ is a  scaling factor and $\alpha\geq 0$ is an exponent. Note that under (\ref{generalstoch}) the probability of selecting a particular restaurant increases with its rank  and decreases with its  popularity in the immediate past (given by the number $n_k(t-1)$). Certain properties of the strategies given by (\ref{generalstoch}) are the following:
\begin{enumerate}
\item For $\alpha=0$ and $T\rightarrow \infty$, $p_k(t)=\frac{1}{N}$ corresponds to the complete random choice case for which we know \cite{anindya} that the utilization fraction is around $0.63$, that is on an average there is 63\% utilization of the  restaurants (see appendix A).

\item For $\alpha=0$ and $T\rightarrow 0$, the agents avoid those restaurants visited last evening and choose again randomly from the remaining restaurants \cite{anindya}. With appropriate simulation it was shown that the distribution of the fraction $f$ of utilization of the restaurants is Gaussian around  $0.46$ (see subsection 2.2).

\end{enumerate}
\subsection{Rank dependent strategies:}  For any natural number $\alpha$ and $T\rightarrow \infty$, an agent goes to the $k$-\rm{th} ranked restaurant with probability $p_k(t)=k^\alpha/\sum k^\alpha$; which means in the limit  $T\rightarrow \infty$ in (\ref{generalstoch}) gives $p_k(t)=k^\alpha/\sum k^\alpha$. Let us discuss the results for such a strategy here. 

If an agent selects any restaurant with equal probability $p$ then probability that a single restaurant is chosen by $m$ agents is given by
\begin{eqnarray}
\Delta(m) &=& \left( \begin{array}{c}  N \\ m \end{array}\right)
p^m (1-p)^{N -m}.
\end{eqnarray}

\noindent Therefore, the probability that a restaurant with rank $k$ is not chosen by any of the agents will be given by
    
 \begin{eqnarray}
\Delta_k(m=0) &=& \left( \begin{array}{c}  N \\ 0 \end{array}\right)
\left(1-p_{k}\right)^{ N }; \ \ p_k=\frac{k^\alpha}{\sum k^\alpha}  \nonumber \\
&\simeq& \exp\left({-k^\alpha N\over \widetilde {N} }\right) \ \ {\rm as} \ \ N \to \infty,
\end{eqnarray}

\noindent where $\widetilde N=\sum_{k=1}^{N} k^\alpha\simeq\int_0^N k^\alpha dk= \frac{N^{\alpha+1}}{(\alpha+1)}.$ Hence
\begin{equation}
 \Delta_k(m=0)=\exp\left(-{k^\alpha \left(\alpha+1\right)\over N^\alpha}\right) .
\end{equation}

\normalsize
\noindent Therefore the average fraction of agents getting dinner in the $k$-{\rm th} ranked restaurant is given by
\begin{equation}
\bar f_k=1- \Delta_k\left(m=0\right)
\end{equation}

\begin{figure}[h]
\centering
\includegraphics[bb=50 50 410 302]{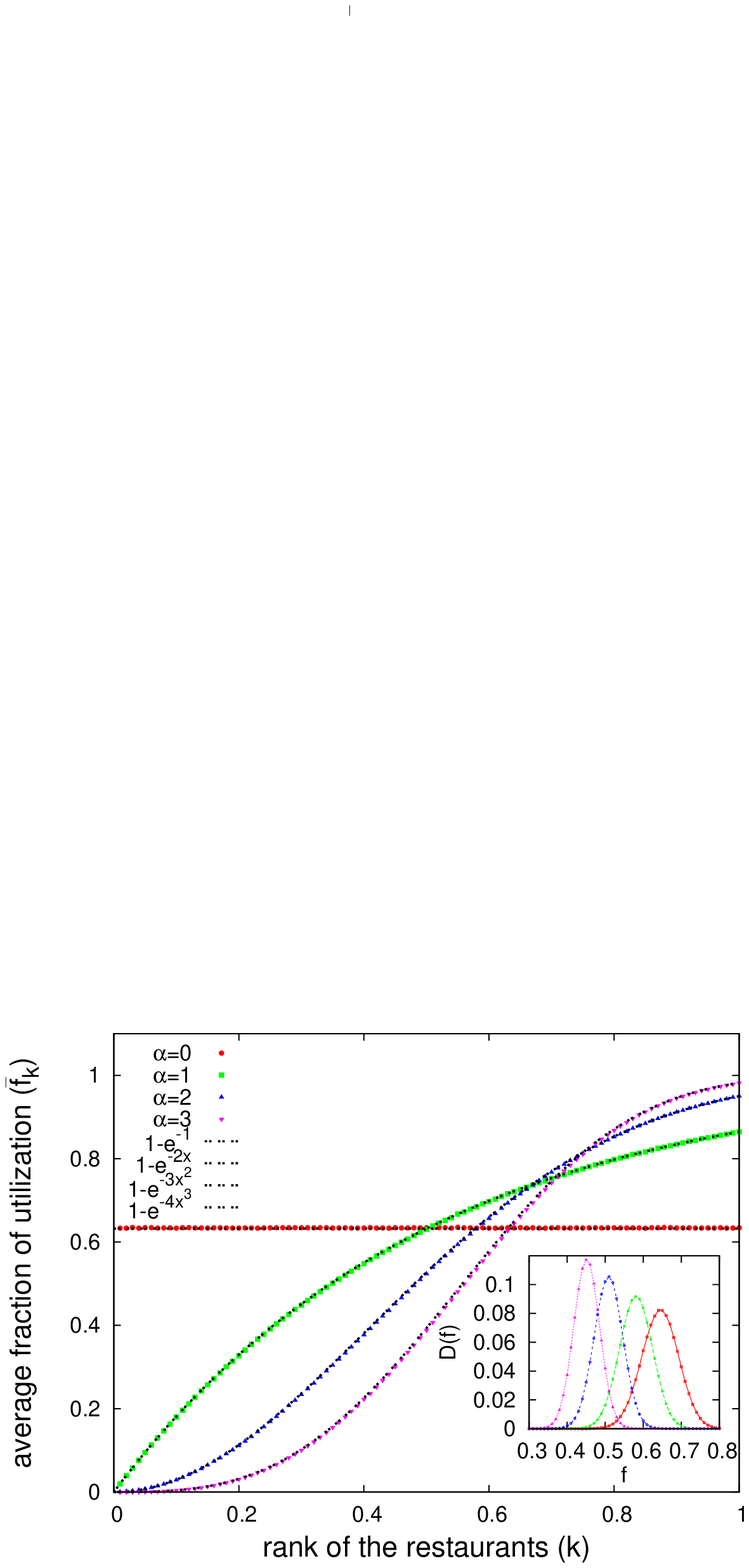}
\caption{The main figure shows average fraction of utilization ($\bar f_k$) versus rank of the restaurants ($k$)  for different $\alpha$ values. The inset shows the distribution $D(f=\sum \bar f_k/N)$ of the fraction $f$ agent getting dinner any evening for different  $\alpha$ values.}
 
\label{fig1}
\end{figure}

\noindent and the  numerical estimates of $\bar f_k$ is shown in Fig. (\ref{fig1}). Naturally for $\alpha=0$, the problem corresponding to random choice $\bar f_k=1-e^{-1}$, giving  $\bar f=\sum \bar f_k/N \simeq 0.63$ and for  $\alpha=1$, $\bar f_k=1-e^{-2k/N}$ giving  $\bar f=\sum \bar f_k/N \simeq 0.58$  as already obtained analytically earlier (see appendix B).

\subsection{Strict crowd-avoiding case}

\noindent We consider here the case (see also \cite{proc})where  each agent chooses on any evening ($t$) randomly among the restaurants in which nobody had gone in the last evening ($t-1$). This correspond to the case where $\alpha =0$ and $T\rightarrow0$ in Eq. (\ref{generalstoch}). Our numerical simulation results for the distribution $D(f)$ of the fraction  $f$ of utilized restaurants is again Gaussian with a most probable  value  at $\bar f\simeq0.46$. This can be explained in the following way: As the fraction $\bar f$ of restaurants  visited by the agents in the  last evening is avoided by the agents this evening, the number of available restaurants is $N(1-\bar f)$ for this evening and is chosen randomly by all the  $N$ agents. Hence, when fitted to Eq.  (\ref{eq:poisson_mm} in  appendix A), $\lambda=1/{(1-\bar f)}$. Therefore, following Eq. (\ref{eq:poisson_mm}), we can write the  equation for $\bar f$ as    

\begin{equation}
(1-\bar {f})\left[1-{\rm exp}\left(-\frac{1}{1-\bar {f}}\right)\right]=\bar {f} .
\end{equation}

\noindent The solution of this equation gives  $\bar f\simeq0.46$. This result agrees well  with the numerical results for this limit ($\alpha=0$, $T\rightarrow0$).

\subsection{Stochastic crowd avoiding case}
\begin{figure}[h]
\centering
 \includegraphics[bb=50 50 410 302]{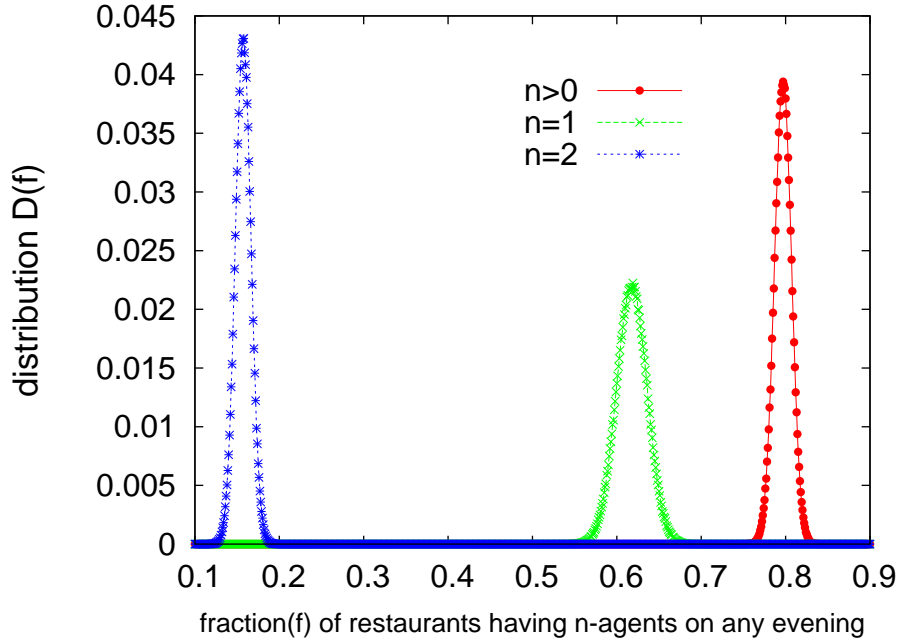}
\caption{Numerical simulation results for a typical prospective customer  distribution on any evening.}
 
\label{fig2}
\end{figure}

In this section we start with the following stochastic strategy: if an agent goes to restaurant $k$ in period ($t-1$) then  the agent goes to the same restaurant in the next period with probability $p_k(t)=\frac{1}{n_k(t-1)}$ and to any other restaurant $k'(\not =k)$ with probability $p_{k'}(t)=\frac{(1-p_k(t))}{(N-1)}$. In this process, the average  utilization fraction is $\bar f\simeq 0.8$ and the distribution $D(f)$ is a Gaussian around $f\simeq0.8$ (see Fig.  \ref{fig2}).

An approximate estimate of $\bar f$: Let $a_i$ denote the fraction  of restaurants where exactly $i$ agents $(i=0,\ldots,N)$ appeared on any evening and assume that $a_i=0$ for $i \geq 3$. Therefore, $a_0+a_1+a_2=1$, $a_1+2a_2=1$ and hence $a_0=a_2$. Given the strategy,   $a_2$ fraction of agents will make  attempts to leave their respective restaurants in the next evening $(t+1)$, while no intrinsic activity will occur on the restaurants where, no body came ($a_0$) or only one came ($a_1$) in the previous evening $(t)$. These $a_2$ fraction of agents will now get equally divided (each in the remaining $N-1$ restaurants). Of these $a_2$, the fraction  going to the vacant restaurants ($a_0$ in the earlier evening) is $a_0a_2$. Hence the new fraction of vacant restaurants is now $a_0-a_0a_2$. In restaurants having exactly two agents ($a_2$ percent in the last evening), some vacancy will be created due to this process, and this is equal to $\frac{a_2}{4}-a_2\frac{a_2}{4}$. Steady state implies that  $a_0-a_0a_2+\frac{a_2}{4}-a_2\frac{a_2}{4}=a_0$ and hence using $a_0=a_2$ we get $a_0=a_2=0.2$,  giving $a_1=0.6$ and $\bar f=a_1+a_2=0.8$. Of course, the above calculation is approximate as none of the restaurant is assumed to get more than two costumers on any evening ($a_i=0$ for $i\geq 3$). The advantage in assuming $a_0$, $a_1$ and $a_2$ only to be non vanishing on any evening is that the activity of redistribution  on  the next evening starts from this $a_2$ fraction of the restaurants. This of course affects $a_0$ and $a_1$ for the next evening   and  for steady state these changes must balance. The computer simulation results also conform that $a_i\leq0.03$ for $i\geq 3$  and hence the above approximation does not lead to  serious error.

\section{Evolving Stochastic Strategy}
\begin{figure}[h]
\centering
 \includegraphics[bb=50 50 410 302]{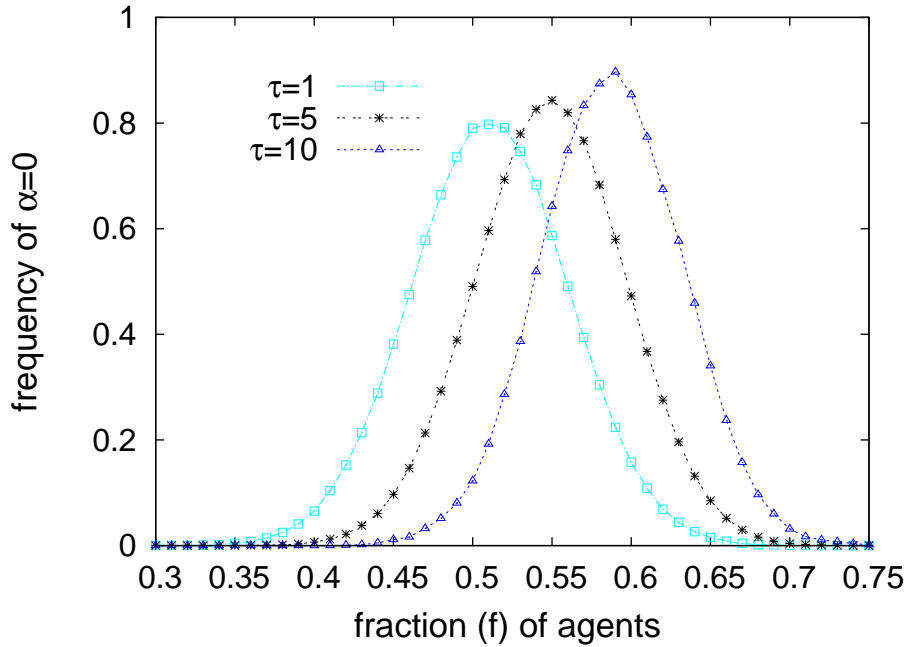}
\caption{Steady state distribution of successful agents having $\alpha=0$. The same for $\alpha=1$ will be given by just complementary function.}
\label{fig3}

\end{figure}

In this section we assume that agents have two possible exogenously given values of  $\alpha$:  $\alpha=0$ or $\alpha=1$. We start by taking some random allocation of $\alpha$ over the set of $N$ agents. The strategy followed by each agent thereafter is the following: if an agent starts with an $\alpha=0 (1)$ and fails to get dinner for the successive $\tau$ evenings then, in the next evening , the agent shifts to $\alpha=1(0)$. The steady state distribution of the $\alpha$ values in the population of agents do not depend on the initial allocation of $\alpha$  values in the population (see Fig. \ref{fig3}). However, as in obvious, for large values of $\tau\geq N$, the stability of the distribution disappears.

\section{Convergence to a fair social norm with deterministic strategies}
In the KPR problem if the rational agents interact then a {\it fair social norm} that can evolve is a periodically organized state with periodicity $N$ where each agent in turn gets served in all the $N$ restaurants and all agents get served every evening. Can we find deterministic strategies (in the absence of a dictator) such that the society achieves this fair social norm?  There is one variant of Pavlov's win shift lose stay strategy (see \cite{kandori,Nowak,orlean}) that can be adopted to achieve the fair social norm and another variant that can be adopted to achieve the fair social norm in an asymptotic sense. Of course, these strategies are deterministic in nature.
\subsection{Fair strategy}
The fair strategy works as follows:
\begin{enumerate}
\item[(i)] At time (evening) $t=0$, agents can choose any restaurants either randomly or deterministically.
\item[(ii)] If at time  $t$ agent $i$ was in a restaurant ranked $k$ and was served then, at time  $t+1$, the agent moves to the restaurant ranked $k-1$ if $k>1$ and moves to the restaurant ranked $N$ if $k = 1$.
\item[(iii)] If agent $i$ was in a restaurant ranked $k$ at time  $t$ and was not served then, at time  $t + 1$, the agent goes to the same restaurant.
\end{enumerate}
It is easy to verify that this strategy gives a convergence to the fair  social norm in less than or equal to $N$ periods. Moreover, after convergence is achieved, the fair social norm is retained ever after. The difficulty with this strategy is that a myopic agent will find it hard to justify the action of going to the restaurant ranked last after getting served in the best ranked restaurant. However, if the agent is not that myopic and observes the past history of strategies played by all the agents and can figure out that this one evening  loss is a tacit commitment devise for this kind of symmetric strategies to work then this voluntary loss is not that implausible. Therefore one needs to run experiments before arguing for or against this kind of symmetric deterministic strategies. More importantly the fair strategy can be modified to take care of this justification problem provided one wants to achieve the fair social norm in an asymptotic sense.
\subsection{Asymptotically fair strategy} The asymptotically fair strategy works as follows:
\begin{enumerate}
\item[(i)] At time (evening) $t=0$, agents can choose any restaurants either randomly or deterministically.
\item[(ii)] If at time $t$ agent $i$ was in a restaurant ranked $k$ and was served then, at time  $t + 1$, the agent moves to the restaurant ranked $k-1$ if $k>1$ and goes to the same restaurant if $k=1$.
\item[(iii)] If agent $i$ was in a restaurant ranked $k$ at time $t$ and was not served then, at time  $t+1$, the agent goes to the restaurant ranked $N$.

\end{enumerate}

\section{Summary and Discussion} 
\noindent We consider the KPR problem where the decision made by each agent in each time period $t$ is independent and is based on the information about the rank $k$ of the restaurants and their occupancy given by the numbers $n_k(t-1)\ldots n_k(0)$. We consider here in Sec. $2$ several stochastic strategies where  each agent chooses the $k$-{\rm th} ranked restaurant with probability $p_k(t)$ given by Eq. (\ref{generalstoch}). The utilization fraction $f_k$ of the $k$-{\rm th} ranked  restaurants on every evening is studied and their  average (over $k$) distributions $D(f)$ are shown in Fig. \ref{fig1} for some special cases. From numerical studies, we find their distributions to be Gaussian with the most probable utilization fraction $\bar f\simeq 0.63$, $0.58$ and $0.46$ for the cases with $\alpha=0$, $T\rightarrow\infty$; $\alpha=1$, $T\rightarrow\infty$; and $\alpha=0$, $T\rightarrow0$ respectively. For the stochastic crowd-avoiding strategy discussed on Sec. $2.3$, we get the best utilization fraction $\bar f\simeq0.8$. The analytical estimates for $\bar f$ in these limits are also given and they agree very well with the numerical observations.

 Finally, we suggest ways to achieve the fair social norm either exactly in the presence of incentive problem or asymptotically in the absence of such incentive problem. Implementing or achieving such a norm in a decentralized way is impossible when $N\rightarrow\infty$ limit. The KPR problem has similarity with the Minority Game Problem \cite{challet2005} as in both the games, herding behavior is punished and diversity's encouraged. Also, both involves learning of the agents from the past successes etc. Of course, KPR has some simple exact solution limits, a few of which are discussed here. In none of these cases considered here, learning strategies are individualistic; rather all the agents choose following the probability given by Eq. (\ref{generalstoch}). In a few different limits of such a learning strategy, the average utilization fraction $\bar f$ and their distributions are obtained and compared with the analytic estimates, which are reasonably close. Needless to mention, the real challenge is to design algorithms of learning mixed strategies (e.g., from the pool discussed here) by the agents so that the fair social norm emerges eventually even when every one decides on the basis of their own information independently. As we have seen,      some naive strategies give better values of $\bar f$ compared to most of the ``smarter'' strategies like strict crowd-avoiding strategies (sec $2.2$) etc. This observation in fact compares well with earlier observation in minority games (see e.g., \cite{Sornette}).

It may be noted that all the stochastic strategies, being  parallel in computational mode, have the advantage that they converge to solution at smaller time steps ($\sim\surd N$ or weakly dependent on $N$) while for deterministic strategies the convergence time is typically of order of $N$, which renders such strategies useless in the truly macroscopic ($N\rightarrow\infty$) limits. However, deterministic strategies are useful when $N$ is small and rational agents can design appropriate punishment schemes for the deviators (see \cite{kandori}).
 
In brief, the study of the KPR problem shows that while a dictated solution leads to one of the best possible solution to the problem, with each agent getting his dinner at the best ranked restaurant with a period of $N$ evenings,  and  with best possible value of $\bar{f}$ ($=1$) starting from the first evening itself. The parallel decision strategies (employing evolving algorithms by the agents, and past informations,   e.g.,  of $n(t)$), which are necessarily parallel among the agents and stochastic  (as in democracy), are less efficient ($\bar {f}\ll1$; the best one discussed here in sec. $2.3$,  giving $\bar {f}\simeq0.8$ only).  We  also note that most of the  ``smarter'' strategies lead to much lower efficiency. 

Is there an upper bound for the value of utilization fraction $\bar f$ (less than unity;  easily achieved in the dictated solution) for such stochastic strategies employed in parallel (democratically) by the agents in KPR? If so, what is this upper bound value? Also, what is the learning time required to arrive at such a solution (compared to zero waiting time to arriving at the most efficient dictated solution) in KPR? These are the questions are to be investigated in future.

\ack
The authors would like to thank Anindya Sundar Chakrabarti and Satya Ranjan Chakravarty for useful comments and discussions.

\appendix
\section{Random-choice case}

\noindent  Suppose there are $\lambda N$ agents and $N$ restaurants. An agents can select  any restaurant with equal probability. Therefore, the probability that a single restaurant is chosen by $m$ agents is given by a Poission distribution in the limit $N\rightarrow\infty$:
\begin{eqnarray}
\label{eq:poisson_mm}
\Delta(m) &=& \left( \begin{array}{c} \lambda N \\ m \end{array}\right)
p^m (1-p)^{\lambda N -m}; \ \ p=\frac{1}{N}  \nonumber \\
&=& \frac{\lambda^m}{m!} \exp({-\lambda}) \ \ {\rm as} \ \ N \to \infty.
\end{eqnarray}
 Therefore the fraction of restaurants not chosen by any agents is given by  $\Delta(m=0) = \exp(-\lambda)$ and that  implies that average fraction of restaurants occupied on any evening is given by \cite{anindya}

 \begin{equation}
\label{rc2}
\bar{f}= 1- \exp(-\lambda) \simeq 0.63 \ {\rm for} \ \lambda=1,
\end{equation}
in the KPR problem. 

\section{Strict-rank-dependent choice }

\noindent In this case, an agent goes to the $k$-\rm{th} ranked restaurant with probability $p_k(t)=k/\sum k$; that is, $p_k(t)$  given by (\ref{generalstoch}) in the limit $\alpha=1$, $T\rightarrow \infty$. Starting with  $N$ restaurants and $N$ agents, we make $N/2$ pairs of restaurants and  each pair has restaurants ranked $k$ and $N+1-k$ where $1\leq k \leq N/2$. Therefore, an agent chooses any pair of restaurant with uniform probability $p=2/N$ or $N$ agents chooses randomly from $N/2$ pairs of restaurants. Therefore the fraction of pairs selected by the agents (from Eq. ~(\ref{eq:poisson_mm}))

 \begin{equation}
f_0= 1- \exp(-\lambda) \simeq 0.86 \ {\rm for} \ \lambda=2.
\end{equation}

\noindent Also, the expected number of restaurants occupied in a pair of restaurants  with rank $k$ and $N+1-k$
 by a pair of agents is
\begin{equation}
E_{k}=1\times\frac{k^2}{(N+1)^2}+1\times\frac{(N+1-k)^2}{(N+1)^2}+2\times2\times\frac{k(N+1-k)}{(N+1)^2}.
\end{equation}
Therefore, the fraction of restaurants  occupied by pairs of agents
\begin{equation}
 f_1=\frac{1}{N}\sum_{k=1,...,N/2} E_{k}  \simeq 0.67.
\end{equation}
Hence, the actual fraction of restaurants occupied by the agents is

 \begin{equation}
\label{rank}
\bar f=f_0.f_1\simeq0.58.
\end{equation}

\end{document}